\documentclass[pra,twocolumn,showpacs,aps,floatfix,amsmath,amssymb,amsfonts]{revtex4}

\usepackage{amsmath}
\usepackage{graphicx}
\usepackage{color}

\newcommand{\mbf}{\mathbf}
\newcommand{\mrm}{\mathrm}

\begin{document}
\title{Reactive collisions of two ultracold particles in a harmonic trap}
\author{Joanna Jankowska$^{1}$ and Zbigniew Idziaszek$^{2}$}
\affiliation{$^1$ College of Inter-Faculty Individual Studies in Mathematics and Natural Sciences,  University of Warsaw, {\.Z}wirki i Wigury 93, 02-089 Warsaw, Poland \\  $^2$ Faculty of Physics, University of Warsaw, Ho{\.z}a 69,
00-681 Warsaw, Poland}
\pacs{34.50.Cx,34.50.Lf}

\date{\today}

\begin{abstract}

We consider two ultracold particles confined in spherically symmetric harmonic trap and interacting via isotropic potential with absorbing boundary conditions at short range that models reactive scattering. First, we apply the contact pseudopotential with complex scattering length and investigate the properties of eigenergies
and eigenfunctions for different trap states as a function of real and imaginary part of the scattering length. In the analyzed case the eigenenergies take complex values, and their imaginary part can be interpreted in terms of decay rates. Then we introduce the model of a square well with absorbing boundary conditions that is later used to investigate the effects of the finite range of interaction on the properties of a reactive scattering in the trap. Finally, we analyze the decay rates for some reactive alkali dimer molecules assuming short-range probability of reaction equal to unity. In this case the complex scattering length is given by universal value related to the mean scattering length of van der Waals interaction between molecules.

\end{abstract}

\maketitle

\section{Introduction}

Ultracold gases of molecules attract an increasing experimental and theoretical interest \cite{Carr2009,PSJ2012}.
Some of them undergo chemical reactions that can be controlled by external fields \cite{Ni2010}, by the internal spin state \cite{Ospelkaus2010} or by aligning dipolar molecules in optical lattice structures of reduced dimensions
\cite{Micheli2010,Quemener2010,Quemener2011,Julienne2011,Zhu2013,simoni2015}. Understanding reactive processes at extremely low temperature is very relevant in quantum chemistry \cite{Carr2009,PSJ2012}. While reactions of highly reactive species can be well understood with the help of universal models derived from the properties of the long-range potentials \cite{Idziaszek2010,Quemener2010,Idziaszek2010a,Jachymski2013,Jachymski2014}, collisions of nonreactive molecules is far more complicated, as the scattering is affected by the presence of a dense spectrum of overlapping resonances. The latter case can be treated by models combing quantum-defect methods with assumptions derived from random-matrix theory \cite{Mayle2012,Mayle2013}.

In this work, we present an analytic treatment of ultracold reactive collisions between particles interacting with an isotropic potential, such as S-state atoms or rotationless polar molecules in the absence of an external electric field, confined in a three-dimensional harmonic trap.
Our modeling is based on the application of quantum-defect theory and includes some key simplifying assumptions regarding the short range molecular dynamics. In particular, when the short range chemical reactivity is high, molecular collisions exhibit universal elastic and inelastic collision properties that depend only on the long range potential. Our results may also find application to nonreactive alkali dimer molecules in excited vibrational levels, since theoretical calculations \cite{Quemener2005,Quemener2007} suggest that their vibrational levels quench to lower vibrational levels with near-universal rate constants. In this work we focus on scattering in s-wave channel, which dominates at ultracold temperatures in collisions of bosonic or distinguishable molecules. Harmonic trapping of polar molecules can be realized by confining them in wells of an optical lattice, which was recently demonstrated for a number of cases \cite{Danzl2010,yan2013,Hazzard2014,Rb2ground,Covey2015}. In practice, alkali dimer molecules may be formed by STIRAP transfer process from weakly bound Feshbach molecules directly in the optical lattice \cite{Danzl2010,Rb2ground,Covey2015}. Realization of the quantum system with two molecules at a single lattice cell, can be envisioned by trapping the molecules in a periodic structure of a double-well potentials, and later by lowering the barrier separating the wells \cite{Anderlini2007}.

This work is structured as follows. In section II we discuss properties of solutions for two particles interacting via regularized $\delta$- pseudopotential with complex scattering length and confined in isotropic harmonic trap. In section III we present the model of a square well potential with absorbing boundary conditions, which is later used in section IV to discuss the effects of the finite range interactions on reactive collisions in traps. Finally, in Section V we apply our model to bosonic KRb and LiCs dimer molecules and predict they decay rates in tight harmonic traps.
We discuss the results and conclude in Sec. VI.

\section{Contact pseudpotential model for trapped molecules} \label{S:busch}

\subsection{Complex scattering length}

In our investigation we consider collisions of two S-state atoms or two molecules in the ground rovibrational state. We model loss collision or the chemical reaction between them using the Fermi pseudopotential
\begin{equation}
U(\mbf{r}) = \frac{2 \pi \hbar^2 a}{\mu} {\delta}(\mbf{r})\frac{\partial}{\partial{r}}r
\end{equation}
with complex scattering length $a = \alpha - i \beta$ and reduced mass $\mu$. This assumes that characteristic range of van der Waals interaction between particles is much smaller than typical length scale of the trapping potential. The imaginary part $\beta>0$ is responsible for reactive collisions and loss of particles from the entrance channel.
The real and the imaginary part of the scattering length determine elastic ${\cal K}^\mrm{el}$ and reactive ${\cal K}^\mrm{re}$ rate constants of molecules in the ultracold regime
\begin{align}
\mathcal{K}^\mrm{el}(E) & = g \frac{\pi \hbar}{\mu k} \left| 1 - S_{\gamma \gamma}(E) \right|^2 = 2 g \frac{h k}{\mu} |a|^2 f(k) \,,   \label{Kel} \\
\mathcal{K}^\mrm{ls}(E) & = g\frac{\pi \hbar}{\mu k} \left(1- |S_{\gamma \gamma}E)|^2\right) =
2g \frac{h}{\mu} \beta f(k) \,, \label{Kloss}
\end{align}
where $k =2 \mu (E-E_\gamma)/\hbar^2$ with $E$ denoting the total energy, $S_{\gamma \gamma}(E)$ is the diagonal element of $S$ matrix corresponding to the entrance channel $\gamma$, and $E_\gamma$ the threshold energy of the channel $\gamma$. The factor $g=1$ except that $g=2$ when both particles are identical species in identical internal states, and partial wave quantum number $\ell$ is restricted to being even (odd) in the case of identical bosons (fermions). The function
\begin{equation}
\label{fellm}
f(k) = \frac{1}{1+k^2|a|^2+2 k \beta},
\end{equation}
has the property that $0< f(k)\le 1$ and $ f(k) \to 1$ as $k \to 0$.  The latter is true when both conditions $k |a| \ll 1$ and $ k \beta \ll 1$ are met.

\subsection{Hamiltonian}

In our model we refer to approach presented in the paper of Busch {\it et al.} \cite{busch}. We consider two identical particles trapped in a harmonic potential and interacting with a~pointlike potential of zero range. The general Hamiltonian for a system consisting of two particles (without internal degrees of freedom) interacting via contact potential in an external isotropic harmonic oscillator trap takes the~form

\begin{equation}\label{eq:hamiltonian_b}
H = -\frac{\hbar^2}{2 m}( {\Delta}_1 + {\Delta}_2 ) + \frac{1}{2}m \omega^2(r_1^2 + r_2^2) + \frac{2 \pi \hbar^2 a}{\mu} {\delta}(\mbf{r})\frac{\partial}{\partial{r}}r   ,
\end{equation}
where $\mbf{r}_1$ and $\mbf{r}_2$ are the~position vectors of the two particles, $\mbf{r}:= {\mbf{r}}_{2}-{\mbf{r}}_{1}$, and $\omega$ is the trap frequency. In the case of harmonic potential, the Hamiltonian can be separated into a center-of-mass part and the relative part, $H = H_{CM}+H_{rel}$. The $H_{CM}$ part takes the simple form of harmonic oscillator Hamiltonian, while the relative-part Hamiltonian  takes form:
\begin{equation}
\label{eq:Hrel}
H_\mrm{rel}=-\frac12 \Delta +\frac12 r^2 + 2 \pi a \delta(\vec{r}) \frac{\partial}{\partial r} r.
\end{equation}
Here, we have applied harmonic oscillator units with the harmonic oscillator length $\sqrt{\hbar / (\mu \omega)}$, and energy $\hbar \omega$. Analytical solutions of Hamiltonian \eqref{eq:Hrel} have been derived by Busch {\it et al.} \cite{busch}. The eigenenergies are given by roots of the following equation
\begin{equation}\label{eq:E_nonabs}
2 \frac{\Gamma (-E/2 + 3/4)}{\Gamma ( - E/2 + 1/4)} = \frac{1}{a} \textrm{  .}
\end{equation}

The wave function for $\ell =0$ and $E \ne \frac{3}{2} + 2n, n \in \mathbb{N}$ reads
\begin{equation}\label{eq:psi_nonabs}
\psi (\vec{r}) = Ae^{-r^2/2} \Gamma \left(-\frac{1}{2}E+\frac{3}{4}\right) U\left(-\frac{1}{2}E+\frac{3}{4}, \frac{3}{2}, r^2\right) \textrm{  ,}
\end{equation}
where $\Gamma$ stands for the Euler Gamma function and $U$ -- for the confluent hypergeometric function. The value of energy $E$ has to be calculated from Eq.~\eqref{eq:E_nonabs} defining the eigenergies. For $\ell \neq 0$ both the eigenenergies and wave functions are given by familiar solutions of the harmonic oscillator problem.

\subsection{Eigenergies and eigenfunctions}

\begin{figure*}
\centering
\includegraphics[width=0.7\linewidth]{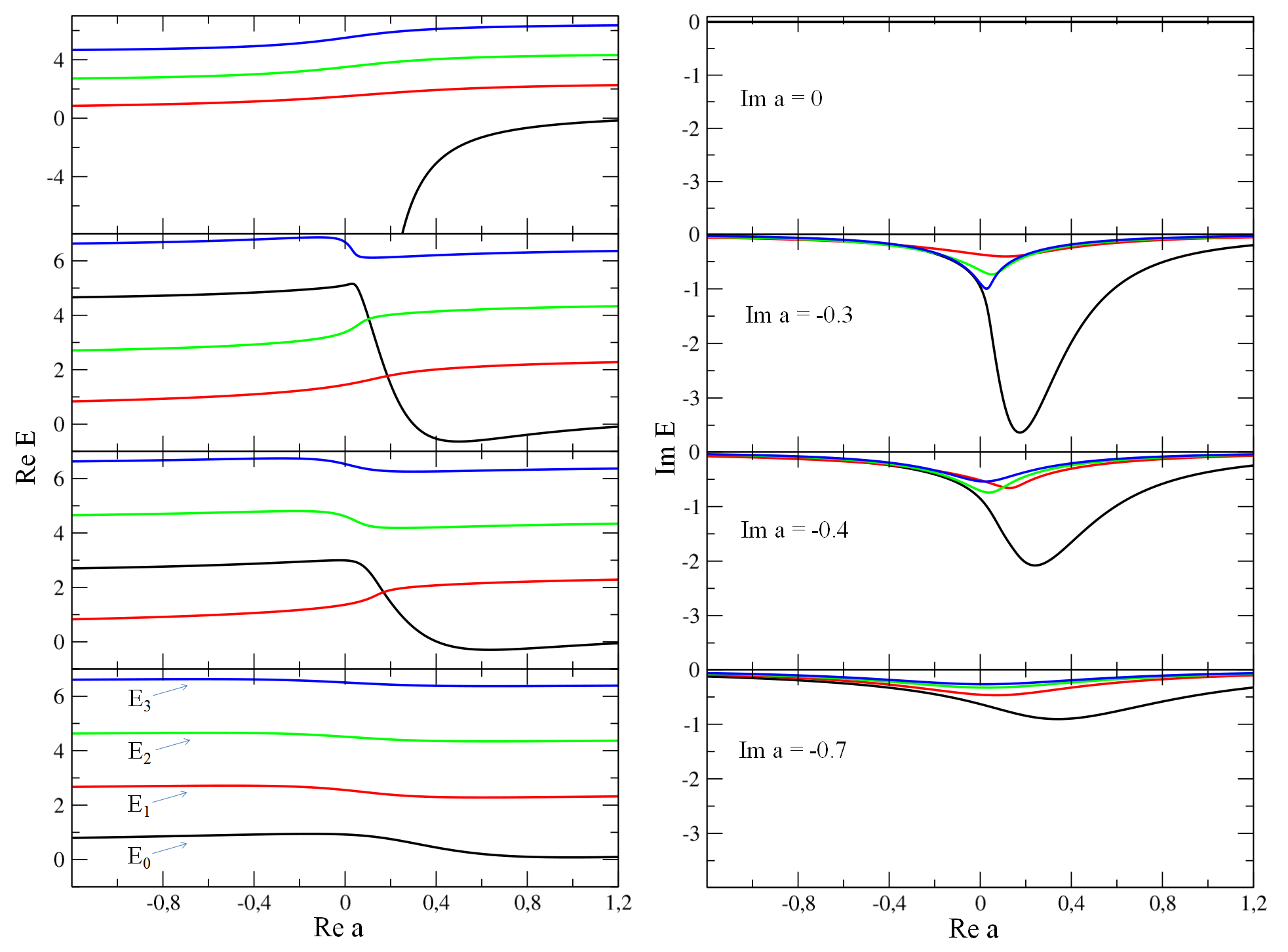}
\caption{$\textrm{Re}\, E$ (left panel) and $\textrm{Im}\, E$ (right panel) dependence on $\textrm{Re}\, a$ for various $\textrm{Im}\, a$ values.}
\label{Fig:busch}
\end{figure*}

As the first step of our analysis we consider the relation $E (\textrm{Re}\, a)$ for fixed $\textrm{Im}\, a$ value (see  Fig.\ref{Fig:busch}). This way we are investigating how the strenght of inelastic interactions affects the system's spectrum for changing elastic part of the relative potential. Left panel of Fig.~\ref{Fig:busch} illustrates the behavior of the real part of the energy for lowest lying energy levels (labeled from $E_0$ to $E_3$ with the increasing real part of the energy at $a \to \infty$ limit). One may see that, in contrast to the nonreactive case ($\textrm{Im}\, a = 0$), for nonzero $\textrm{Im}\, a$ values the energy of the $E_0$ level does not diverge to minus infinity as $a \to 0^+$ anymore. For weak inelastic interaction (small $|\textrm{Im}\, a|$) the energy of the ground state remains negative in some range of $\textrm{Re}\, a$ values but for sufficiently small and positive $\textrm{Re}\, a$ the ground state energy also becomes positive. For small $\textrm{Im}\, a$ we observed that the ground state crosses $E=0$ value for $\textrm{Re}\, a \approx - \textrm{Im}\, a$.

At the same time one may observe the occurrence of level anti-crossings. For small $|\textrm{Im}\, a|$ lowest excited states seem to stay almost unperturbed, while the ground state couples to one of the higher levels. With increasing $|\textrm{Im}\, a|$ value the $E_0$ state couples to lower lying states and finally, for $\textrm{Im}\, a \approx -0.5$ it remains the lowest energy state for all $\textrm{Re}\, a $ values. All levels lying over the coupled state demonstrate a kind of 'reverse' behavior with respect to the pure elastic interacton case -- for more negative $\textrm{Re}\, a$ values their energy is higher than for more positive values.

The right panel of Fig.(\ref{Fig:busch}) shows relation between $\textrm{Im}\, E$ and $\textrm{Re}\, a$. All included levels demonstrate a minimum in imaginary part of the energy. Within investigated range of $\textrm{Im}\, a$ values ($\textrm{Im}\, a \in [0 , -1]$) the minimum of ground state is always the deepest one, however its depth decreases significantly with increasing $|\textrm{Im}\, a|$. The shape of the other states changes to the much lower extent. As will be explained later on, the imaginary part of the energy might be directly related to the lifetime of the molecular complex in particular state. Thus, one may conclude that the ground $E_0$ energy level is especially short-lived. This might be associated with the fact, that in the molecular complex two reactive molecules stay close to each other for most of the time.

\begin{figure*}
\centering
\includegraphics[width=0.95\linewidth]{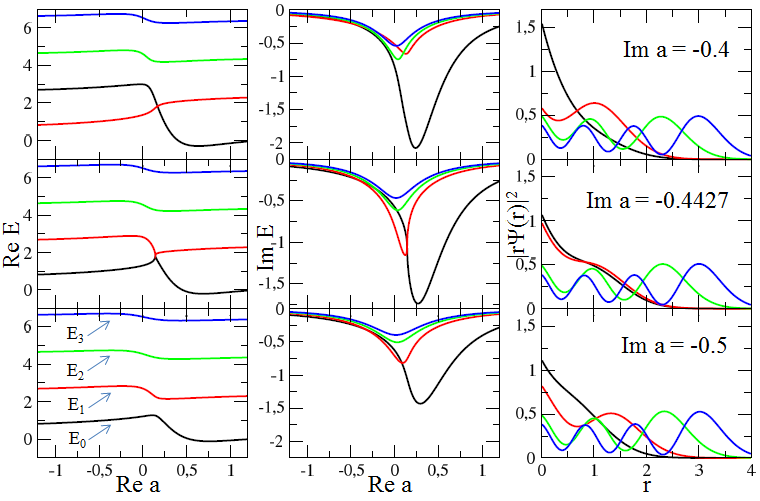}
\caption{$\textrm{Re}\, E$ (left panel), $\textrm{Im}\, E$ (central panel) dependence on $\textrm{Re}\, a$ for various $\textrm{Im}\, a$ values and corresponding radial wave functions  for $ \textrm{Re}\, a$ values selected to minimize the $|E_1-E_0|$. }
\label{Fig:busch_cross_ff}
\end{figure*}

The detailed behavior of the energy levels and the wave function in the regime of parameters close to the characteristic point when the system exhibits an avoided crossing is shown in Fig.~\ref{Fig:busch_cross_ff}. For a reactive system the notion of the avoided crossing has to be extended, as both real and imaginary parts of the energy have to be considered. In Fig.~\ref{Fig:busch_cross_ff} for $\textrm{Im}\, a = 0.4$ levels $E_0$ and $E_1$ seem to cross but their imaginary parts are different. At the same time their wave functions at the point corresponding to the smallest $|E_1-E_0|$ exhibit behavior typical for the ground and first excited state in the trap. For $\textrm{Im}\, a = 0.4427$ the levels exhibit a true avoided crossing as their real and imaginary parts have almost identical values at the crossing point. In addition, their wave functions exhibit similar shape. Finally for $\textrm{Im}\, a = 0.5$ the levels seems to repulse each other in the vicinity of the crossing, and their wave function become again very different.

\subsection{Non-Hermitian Hamiltonian}

Our characterization of the system consisting of two reactively colliding molecules assumes the application of the interaction with complex scattering length. Basic properties of non-Hermitian Schr\"o{}dinger equation are well studied in the literature (see for example  \cite{mohms_czechjp2006,non-Hermitian}). In general, one needs to deal with two sets of eigenvalues ($\{{\lambda}_i\}$ , $\{{\epsilon}_i\}$) and eigenvectors ($\{|{\psi}_i\rangle \}$ , $\{|{\chi}_i\rangle \}$), where $\{{\lambda}_i\}$ and $\{|{\psi}_i\rangle \}$ are right-handed and $\{{\epsilon}_i\}$ and $\{|{\chi}_i\rangle \}$ are left-handed eigenvalues and eigenvectors, respectively. The eigenvalues are in general complex and satisfy the following equations:

\begin{displaymath}
H |{\psi}_i\rangle = {\lambda}_i |{\psi}_i\rangle \quad , \quad  H^{\dagger} |{\chi}_i\rangle = {\epsilon}_i |{\chi}_i\rangle .
\end{displaymath}
Comparison of the equations above leads to the relation between the eigenvalues from the both sets (the eigenspectrum of $H$ is unique):

\begin{displaymath}
 {\lambda}_i = {{\epsilon}_i}^* \quad \textrm{for all $i$.}
\end{displaymath}
The state vectors $|{\psi}_i\rangle $ and $|{\chi}_i\rangle $ are in general different. After a few simple steps, one may find that

\begin{displaymath}
 ({\lambda}_i - {{\epsilon}_j}^*)\langle {\chi}_j|{\psi}_i \rangle = 0.
\end{displaymath}
This allows to choose a generalised orthonormality relationship:

\begin{displaymath}
 \langle {\chi}_j|{\psi}_i \rangle = {\delta}_{ij}.
\end{displaymath}
In order to relate to complex eigenergies with the decay rates of the reactive particles in the trap we consider the time evolution of the density matrix. We start by expressing an arbitrary wave function it in terms of right-hand-side eigenvectors
\begin{equation}
 |{\Psi (t)}\rangle = \sum_{i} c_i(t) |{\psi}_i\rangle \textrm{  ,}
\end{equation}
where $c_i(t)$ are time-dependent expansion coefficients. For a pure state the density operator $\rho(t) = |{\Psi (t)}\rangle \langle {\Psi (t)}|$. In general case

\begin{equation}
\rho (t) = \sum_{i,j} c_{ij}(t) |{\psi}_i\rangle \langle {\psi}_j| \textrm{  .}
\end{equation}
where $c_{ij}(t)$ is the time-dependent coefficient of the expansion. The time evolution of the density matrix is ruled by a generalized Liouville equation (see for example \cite{mohms_czechjp2006, china2003})

\begin{equation}
\dot{\rho} (t) =  \frac{1}{i\hbar} (H\rho(t) - \rho (t) H^{\dagger}) \textrm{  ,}
\end{equation}
Through expansion of the left- and right-hand sides of this equation one arrives at

\begin{equation}
\sum_{i,j} \dot{c}_{ij}(t) |{\psi}_i\rangle \langle {\psi}_j| =  \frac{1}{i\hbar} \sum_{i,j}  c_{ij}(t) ({\lambda}_i |{\psi}_i\rangle \langle {\psi}_j| - |{\psi}_i\rangle \langle {\psi}_j| {\lambda}^*_j) \textrm{  ,}
\end{equation}
Now, projecting both sides of the expression above on left-handed eigenfunctions $\langle {\chi}_i|$ from the left and $|{\chi}_i\rangle$ from the right,
we obtain

\begin{equation}
c_{ij} (t) = c_{ij} (0) \exp \left[\frac{\left({\lambda}_i -{\lambda}^*_j \right) t}{i \hbar} \right] \textrm{  ,}
\end{equation}

The coefficients of the diagonal part of $\rho$ decay according to

\begin{equation}
c_{ii} (t) = c_{ii} (0) e^{2 \textrm{Im}\, {\lambda}_i t/\hbar} \textrm{  ,}
\end{equation}
Taking into account that ${\lambda}_i = {{\epsilon}_i}^*$, the lifetime of each of the right-hand-side eigenstates of $H$ is given by

\begin{equation}
{\tau}_n = -\frac{\hbar}{2 \textrm{Im}\, \epsilon_n} \textrm{  .}
\end{equation}
Thus, the higher modulus of imaginary part, the shorter lifetime of the corresponding state.

\section{Absorbing quantum well}\label{S:AQW}

In the second part of our study we  investigate reactive collisions between ultracold molecules using a model of a square well potential with absorbing boundary conditions at short range. The isotropic potential used in this model allows for description of reactive collisions of particles taking into account some finite range of the interaction potential. Absorbing boundary conditions result in the~particle losses in the~elastic scattering channel. The short range reaction rate is controlled by a single complex parameter $\eta$.

\begin{figure}
\centering
\includegraphics[width=0.7\linewidth]{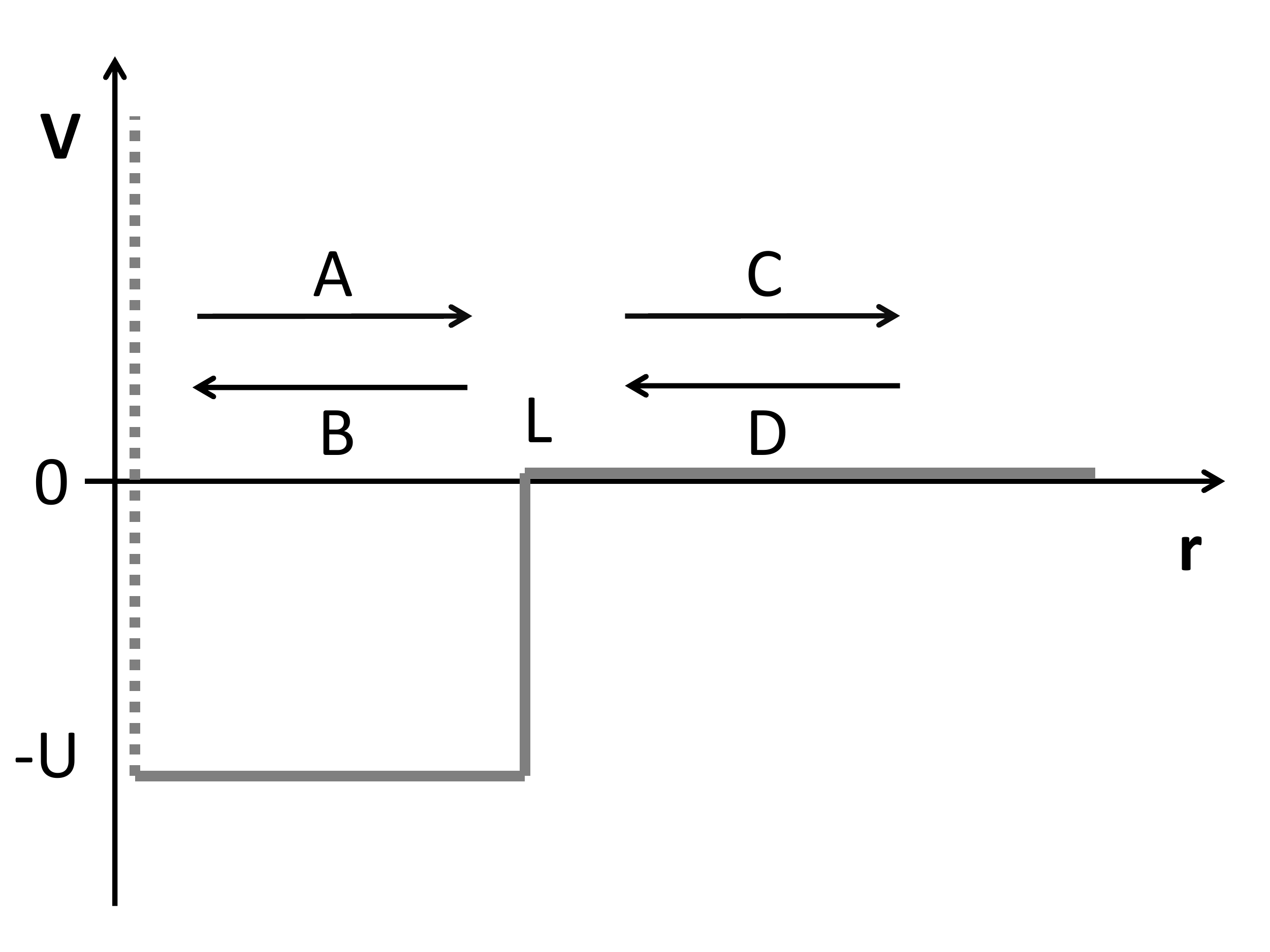}
\caption{Absorbing square quantum well.}
\label{Fig:studnia1}
\end{figure}

Schr\"o{}dinger equation and general forms of the~radial wave functions in the AQW model take the same form as for the well known nonabsorbing case

\begin{equation}
H =- \frac{1}{2}{\Delta}_r + V(r) \textrm{  ,}
\end{equation}
where $r$ parameterizes the radial direction in three-dimensional space\footnote{again, in all cases reduced harmonic oscillator units are used}, while

\begin{equation}
V(r) =
\left\{\begin{array}{c l}
-U  & \textrm{ for $r<L$}\\
0   & \textrm{for $r \ge L$  }
\end{array}\right.
\end{equation}
and

\begin{equation}\label{eq: psi_in}
{\psi}_{in}(r) = A e^{ikr} + B e^{-ikr} \quad \textrm{,  } k := \sqrt{2(E + U)}
\end{equation}
\begin{equation}\label{eq: psi_out}
{\psi}_{out}(r) = C e^{i\kappa r} + D e^{-i \kappa r} \quad \textrm{,  } \kappa := \sqrt{2E} ,
\end{equation}
where ${\psi}_{in}(r) := r \Psi (\vec{r})$ for $r<L$,  ${\psi}_{out}(r) := r \Psi (\vec{r})$ for $r \ge L$ and $A$, $B$, $C$, $D$ are constants.

The~difference with respect to the nonabsorbing system is in the~boundary conditions at short range described by the~new, additional parameter$\eta$: $|\eta | \le 1$, $\eta \in \mathbb{C}$. This parameter modifies the~$A / B$ ratio (which for nonabsorbing well was always equal to unity) and sets the phase of the reflected wave. The requirement of the continuity of the wave function and its derivative at $r = L$:
${\psi}_{in} (L) = {\psi}_{out} (L)$, ${\psi}_{in}^\prime(L) = {\psi}_{out}^\prime(L)$, leads to the following equations
\begin{equation}
\label{Eq:AQW}
\left\{\begin{array}{l}
B \left( e^{-ikL} - \eta e^{ikL}\right) = Ce^{i\kappa L} + De^{-i\kappa L} \\
-ikB\left( e^{-ikL} + \eta e^{ikL}\right) = i\kappa \left (Ce^{i\kappa L} - De^{-i\kappa L}\right) \textrm{  .}
\end{array}\right.
\end{equation}
In the case of absorbing well the standard boundary condition ${\psi}_{in} (0) = A + B =0$
has been replaced with $A +\eta B=0$ . This way $\eta$ parameterizes the short-range density flux of particles. For $\eta=1$ there is no reaction at short range, while for $\eta =0$ the probability of the reaction is equal to unity.

Solving Eqs.~\eqref{Eq:AQW} together with the $A +\eta B=0$ condition, we obtain
\begin{equation}
\tilde{A} = \frac{2\kappa \eta e^{-i\kappa L}}{\kappa \left( \eta e^{ikL} - e^{-ikL}\right) - k\left( \eta e^{ikL} + e^{-ikL}\right)} \textrm{ , }
\end{equation}
\begin{equation}
\tilde{B} = \frac{2\kappa e^{-i\kappa L}}{k\left( \eta e^{ikL} + e^{-ikL}\right) - \kappa \left( \eta e^{ikL} - e^{-ikL}\right)} \textrm{ , }
\end{equation}
\begin{equation}
\tilde{C} = e^{-2i\kappa L}\frac{\kappa \left( \eta e^{ikL} - e^{-ikL}\right) + k\left( \eta e^{ikL} + e^{-ikL}\right)}{\kappa \left( \eta e^{ikL} - e^{-ikL}\right) - k\left( \eta e^{ikL} + e^{-ikL}\right)} ,
\end{equation}
where $\tilde{A}$, $\tilde{B}$ and $\tilde{C}$ means A, B or C divided by arbitrary value of~D, respectively (see also Fig.~\ref{Fig:studnia1}).

One can show that the $s$-wave phase shift $\delta$ can be determined from the~$\tan{\delta} = (\tilde{C}+1)/i(\tilde{C}-1)$ relation. This leads to the explicit expression for the energy-dependent scattering length $a(\kappa) = - \tan \delta (\kappa)/\kappa$,
\begin{widetext}
\begin{equation}
 a(\kappa ) = \frac{i}{\kappa } \left[ 1 + 2\frac{\kappa \left( \eta e^{ikL} - e^{-ikL}\right) - k\left( \eta e^{ikL} + e^{-ikL}\right) }{\kappa \left( \eta e^{ikL} - e^{-ikL}\right) \left( e^{-2i\kappa L} - 1\right) + k\left( \eta e^{ikL} + e^{-ikL}\right) \left( e^{-2i\kappa L} + 1\right)}\right].
\label{Eq:a_kappa}
\end{equation}
\end{widetext}
In the regime of small energies ($E \to 0$) Eq.~\ref{Eq:a_kappa} may be simplified to high extent -- in this case one obtains the expression for energy-independent scattering length which takes takes the form

\begin{equation}
a(\kappa )\to {\tilde{a}}_{\eta}(\alpha) = L \left[ 1 + \frac{i\left( \eta e^{i\alpha } - e^{-i\alpha }\right)}{\alpha \left( \eta e^{i\alpha } + e^{-i\alpha }\right)}\right]
\end{equation}

where $\alpha := \sqrt{2U}L$ .

\begin{figure*}
\centering
\includegraphics[width=0.70\linewidth]{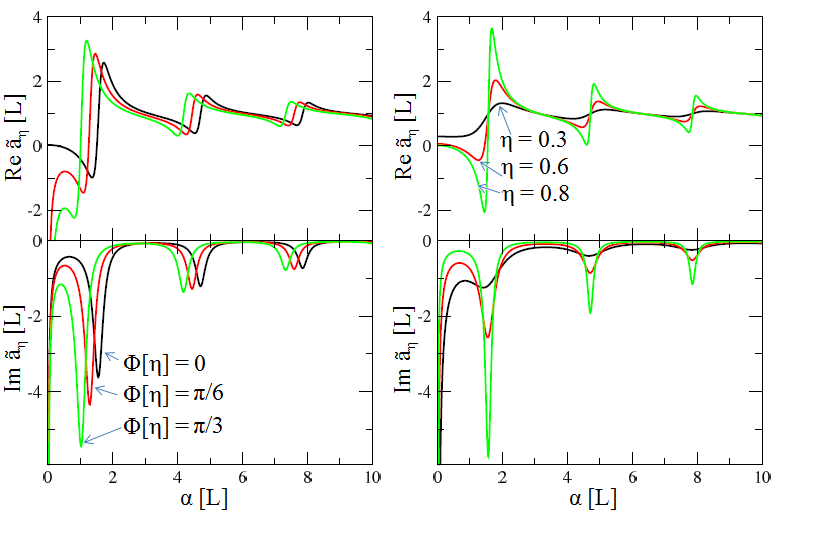}
\caption{$\textrm{Re}\, {\tilde{a}}_{\eta}$ (top) and $\textrm{Im}\, {\tilde{a}}_{\eta}$ (bottom) dependence on $\alpha$ for various $\eta$ phases (left panel, $|\eta| = 0.7$) and for various $\eta$ moduli (right panel,$\Phi [\eta ] = 0$).}
\label{Fig:studnia_blok}
\end{figure*}

 In Fig.(\ref{Fig:studnia_blok}) the dependence of the scattering length on $\alpha$ for various $\eta$ values is presented --  the left panel illustrates dependences for various $\eta$ phases ($\Phi [\eta ]$), while the right panel shows the function for different values of $\eta$ modulus ($|\eta|$). One can see that for $|\eta| \to 1$ the~real part of the scattering length becomes sharper -- more similar to the nonabsorbing case in which the function exhibits discontinuities. These situations correspond to shape resonances of the quantum
well. The~behavior of the imaginary part also changes significantly with the modulus value of the absorption parameter. For small $|\eta|$ values (strong short-range absorption) $\textrm{Im}\, ({\tilde{a}}_{\eta} (\alpha ))$ oscillates in quite regular way (its value is always negative). When $|\eta| \to 1$ again the function is getting sharp. The changes observed for real and imaginary parts of ${\tilde{a}}_{\eta}$ presented in the left panel are of the same type. One may notice that the increase of the phase results in the shift of the peaks' positions toward smaller $\alpha$ values -- this shift is accompanied by the increase in hight/depth. For $\Phi [\eta ] \to 2\pi $ each peak is shifted to the initial position of its neighbor from the side of smaller $\alpha$.

\section{Finite-range interactions of trapped molecules}

\begin{figure}
\centering
\includegraphics[width=0.7\linewidth]{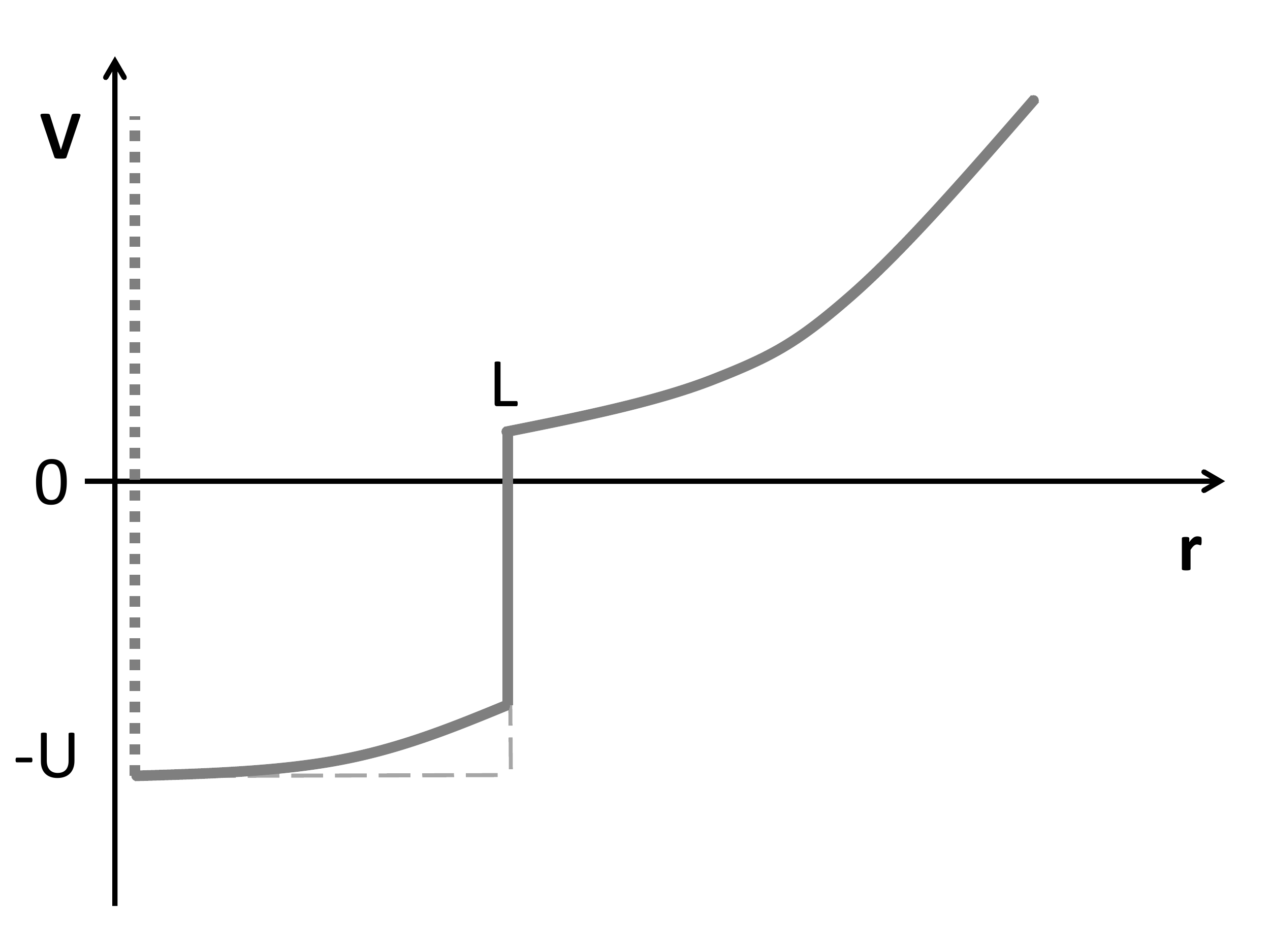}
\caption{Finite-range interaction potential between molecules in a harmonic trap.}
\label{Fig:studnia2}
\end{figure}

In this approach we combine the harmonic trapping potential with the interparticle interaction represented by AQW model introduced in Section \ref{S:AQW} to investigate the finite-range interactions of trapped molecules. The relative Hamiltonian for such system (schematically shown in Fig. \ref{Fig:studnia2}) takes the form

\begin{equation}
H =- \frac{1}{2}{\Delta}_r + \frac{1}{2}r^2 + V(r) \textrm{  , where}
\end{equation}

\begin{equation}
V(r) =
\left\{\begin{array}{c l}
- U  & \textrm{ for $r<L$}\\
0   & \textrm{for $r \ge L$  . }
\end{array}\right.
\end{equation}
The solution of this eigenequation involves some steps -- finally one obtains it in simplified form

\begin{equation}
y\frac{{\partial}^2}{\partial y^2}h(y)+(\ell+\frac{3}{2}-y)\frac{\partial}{\partial y}h(y)+\frac{2\varepsilon -3-2\ell }{4}h(y)=0
\label{eq:konf}
\end{equation}
where $y=r^2$,  $\varepsilon = E+U$ for $r<L$ and  $\varepsilon = E$ for $r>L$, $l$ denotes the partial wave number and $h(y)$ is the function closely related to the wave function of the system, namely $\psi (\vec{r}) = Y_{\ell m}(\theta ,\phi ) r^\ell e^{-r^2/2}h(r^2)$. Eq. \eqref{eq:konf} is known as confluent hypergeometric equation. Its eigenfunctions may be generally expressed as linear combinations of $F[a,b,z]$ -- Kummer's (confluent hypergeometric) function and $U[a,b,z]$ -- Tricomi's (confluent hypergeometric) function. The imposition of absorbing boundary condition consistent with the AQW approach leads to the following form of the radial part $h(y)$ of the wave function

\begin{widetext}
\begin{equation}\label{Eq:hy}
h(y) =
\left\{\begin{array}{l r}
C_1 F\left[ \frac{3+2l-2(E+U)}{4}, \frac{3}{2}+l,y \right]+C_2U \left[ \frac{3+2l-2(E+U)}{4}, \frac{3}{2}+l,y \right] & \textrm{for $r<L$}\\
C_3 U\left[ \frac{3+2l-2E}{4}, \frac{3}{2}+l,y \right] & \textrm{for $r \ge L$ .}
\end{array}\right.
\end{equation}
\end{widetext}
Here, $C$ constants depends on $\eta$ parameter. The eigenvalues $E$ and $C_1/C_2$ fraction may be obtained by imposing of condition for continuity of the logarithmic derivative at $r=L$. For simplicity we also assume that the quantum well is sufficiently deep, such that $L^2 \ll U$ and $|E| \ll U$.

\begin{figure*}
\centering
\includegraphics[width=0.750\linewidth]{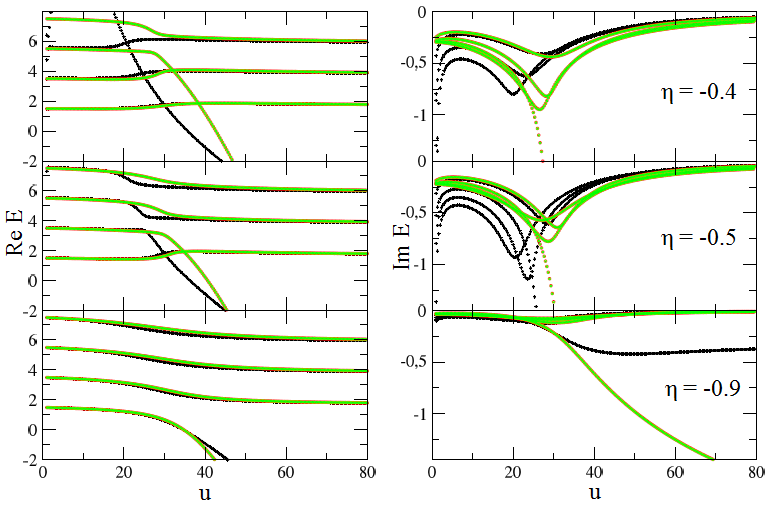}
\caption{$\textrm{Re}\, E$ (left panel) and $\textrm{Im}\, E$ (right panel) dependence on $u$ for various $\eta$ values: black -- the zero-range interaction model with energy independent scattering length, red --  the zero-range interaction model  with energy dependent scattering length, green -- the finite-range interaction approach.}
\label{Fig:jj_&_zi}
\end{figure*}

In our study we solve the eigenproblem presented above for $\ell=0$ case. Fig. \ref{Fig:jj_&_zi} shows the calculated  $E (U)$ dependences for various $\eta$ values (left and right panels illustrate $\textrm{Re}\, E$ and $\textrm{Im}\, E$, respectively). It compares three different approaches to the problem. First of all, the 'exact' solution of the finite-range interaction model is shown. Secondly, the $\delta$ - like contact potential model combined with energy independent scattering length is presented. Finally, we introduce the combination of  $\delta$ - like contact potential model and energy dependent scattering length.

The second approach based on the standard pseudopotential, one hand provides the simplest possible way to solve the problem but on the other, as one may see in Fig. \ref{Fig:jj_&_zi}, fails to reproduce results of the 'exact' solution in some cases (especially close to the avoided-crossing points). The last of the three proposed models links some of the advantages of previous both. Namely, it shows simplicity comparable to the second method but at the same time the correction of energy-dependency applied to the scattering length leads to results almost identical with those obtained with use of the 'exact' approach. Thus, we believe that the it can be considered as a method of choice for this sort of investigation.

\section{Lifetime of molecules in traps}

To apply our results to particular chemical system, we calculated lifetime values for lowest energy levels of two pairs of identical bosonic molecules trapped in a harmonic potential: $^{41}$K$^{87}$Rb and $^{7}$Li$^{133}$Cs dimers. Such molecules have been produced recently in the rovibrational ground state using photoassociation methods \cite{Aikawa2010,Deiglmayr2008}.
In this case we assumed that the initial state is a pure quantum state of a harmonic potential. We modeled the interaction with contact $\delta$ pseudopotential with energy independent scattering length within the universal limit given by $a = \bar{a} - i \bar{a}$ \cite{Idziaszek2010}, assuming the molecules are highly reactive. In this regime the energy independent model may be considered reliable, as long as excitation energy in the trap is much smaller than characteristic energy of van der Waals interaction $\hbar \omega \ll E_\mrm{vdW}$. When the collision energy becomes comparable to $E_\mrm{vdW}$ one expects some energy-dependent corrections to reactive rates calculated in the zero-energy limit \cite{Jachymski2013,Jachymski2014}.

\begin{figure*}
\centering
\includegraphics[width=0.60\linewidth]{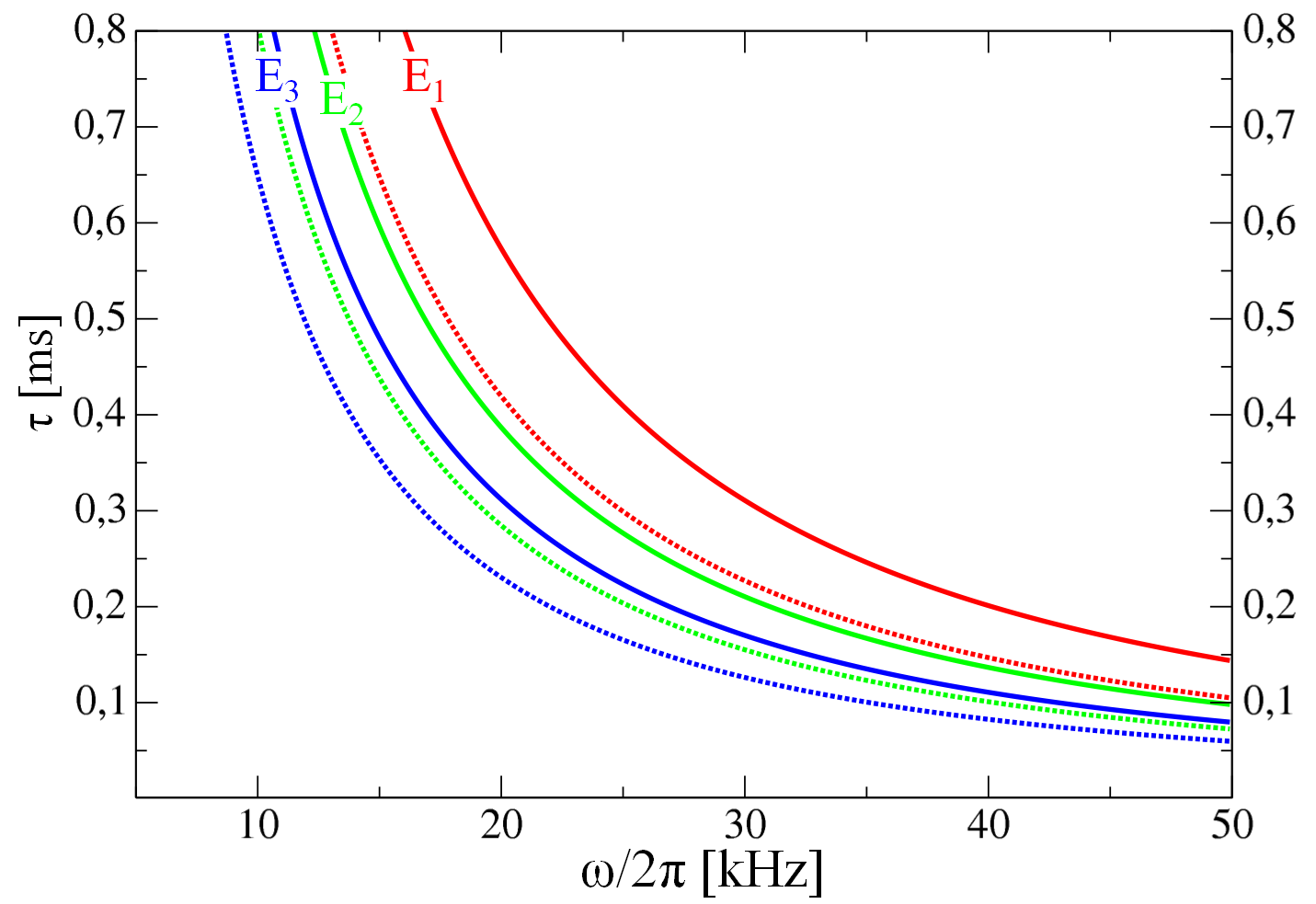}
\caption{Lifetime dependence on harmonic trap frequency for two colliding $^{41}$K$^{87}$Rb (solid lines) and $^{7}$Li$^{133}$Cs (dashed lines) molecules in the universal limit.}
\label{Fig:krb}
\end{figure*}

In Fig. \ref{Fig:krb} the system's lifetime dependence on harmonic trap frequency for three lowest-lying excited energy levels is presented. One may observe, that lifetime strongly depends on the trapping frequency. By increasing the trapping frequency the lifetime of all shown excited states decreases like $\sim1/\omega$. At the same time the ground state lifetime remains well below a microsecond (not shown in Fig. \ref{Fig:krb}). This might be related to the fact that for positive real part of the scattering length, the ground state of two particles in a harmonic trap correspond to a molecular complex, where two molecules stays close to each other. In such a case the reactive loss process is strongly enhanced.

For typical depths of the optical lattice potential, which is of the order of 50 $E_{r}$ (recoil energies) in current experiments \cite{yan2013,Covey2015}, the trapping frequency for KRb molecule at the bottom of the single well is about 20~kHz. For this trapping frequency our model predicts the lifetime of the order of 0.5~ms for KRb and 0.4~ms for LiCs particles in the lowest excited state of a harmonic well. Thus, our model remains compatible with predictions based on the 3D universal rate $K^\mrm{re}=2 \bar{a} \hbar/\mu \approx 1.3 \times 10^{-11}$ applied to the size of the ground state of the harmonic potential, which at 20~kHz is $l =2 a_{HO} = 2 \hbar/{\mu \omega} \approx  180$~nm leading to the lifetime $\tau = 0.5$~ms. In recent experiments, however, the two-body loss rate for collisions of KRb molecules in two different rotational states was measured to be $9.0(4) \times 10^{−10}$~cm$^3$/s \cite{yan2013}, which is much larger than universal two-body loss rate $K^\mrm{re}$. Therefore lifetimes predicted by our theory are about two orders of magnitude larger than the on-site loss rates that were measured in experiments \cite{yan2013}. This suggests that loss rates for collisions of molecules in two different rotational states are probably non-universal.

\section{Conclusions}

In conclusion, we studied the problem of two ultracold particles confined in spherically symmetric harmonic trap and interacting via isotropic potential with absorbing boundary conditions at short range that represents reactive scattering. This can correspond, for instance, to collisions of two molecules in the rovibrational ground state. We applied the contact pseudopotential with complex scattering length and investigate the properties of eigenergies
and eigenfunctions for different trap states as a function of real and imaginary part of the scattering length. In the analyzed case the eigenenergies take complex values, and their imaginary part can be interpreted in terms of decay rates. When increasing the imaginary part of the scattering length the energy spectrum exhibits series of crossings. At the same time the lifetime of a bound state becomes very short and bound state can exist only for sufficiently large real part of the scattering length. In the second part of the paper we introduced the model of a square well with absorbing boundary conditions that was later used to investigate the effects of the finite range of interaction on the properties of a reactive scattering in the trap. We concluded that the real part of the energy spectrum is well reproduced by the contact pseudopotenial apart from the region where the scattering length diverges. The proper description of the imaginary part of the spectrum requires, however, application of the energy-dependent scattering length in the pseudopotential. Finally, we analyzed the decay rates for some reactive alkali dimer molecules assuming short-range probability of reaction equal to unity, for different eigenstates of particles in the trap. In this case the complex scattering length is given by universal value related to the mean scattering length of van der Waals interaction between molecules.

\section{Acknowledgements}

This work was supported by the National Center for Science grant number DEC-2011/01/B/ST2/02030.

\bibliography{jjankowska_zidziaszek}

\begin{thebibliography}{30}
\expandafter\ifx\csname natexlab\endcsname\relax\def\natexlab#1{#1}\fi
\expandafter\ifx\csname bibnamefont\endcsname\relax
  \def\bibnamefont#1{#1}\fi
\expandafter\ifx\csname bibfnamefont\endcsname\relax
  \def\bibfnamefont#1{#1}\fi
\expandafter\ifx\csname citenamefont\endcsname\relax
  \def\citenamefont#1{#1}\fi
\expandafter\ifx\csname url\endcsname\relax
  \def\url#1{\texttt{#1}}\fi
\expandafter\ifx\csname urlprefix\endcsname\relax\def\urlprefix{URL }\fi
\providecommand{\bibinfo}[2]{#2}
\providecommand{\eprint}[2][]{\url{#2}}

\bibitem[{\citenamefont{Carr et~al.}(2009)\citenamefont{Carr, DeMille, Krems,
  and Ye}}]{Carr2009}
\bibinfo{author}{\bibfnamefont{L.~D.} \bibnamefont{Carr}},
  \bibinfo{author}{\bibfnamefont{D.}~\bibnamefont{DeMille}},
  \bibinfo{author}{\bibfnamefont{R.~V.} \bibnamefont{Krems}}, \bibnamefont{and}
  \bibinfo{author}{\bibfnamefont{J.}~\bibnamefont{Ye}}, \bibinfo{journal}{New
  Journ. Phys.} \textbf{\bibinfo{volume}{11}}, \bibinfo{pages}{055049}
  (\bibinfo{year}{2009}).

\bibitem[{\citenamefont{Qu\`{e}m\`{e}ner and Julienne}(2012)}]{PSJ2012}
\bibinfo{author}{\bibfnamefont{G.}~\bibnamefont{Qu\`{e}m\`{e}ner}}
  \bibnamefont{and} \bibinfo{author}{\bibfnamefont{P.~S.}
  \bibnamefont{Julienne}}, \bibinfo{journal}{Chemical Reviews}
  \textbf{\bibinfo{volume}{112}}, \bibinfo{pages}{4949} (\bibinfo{year}{2012}).

\bibitem[{\citenamefont{Ni et~al.}(2010)\citenamefont{Ni, Ospelkaus, Wang,
  Qu{\'e}m{\'e}ner, Neyenhuis, de~Miranda, Bohn, Ye, and Jin}}]{Ni2010}
\bibinfo{author}{\bibfnamefont{K.-K.} \bibnamefont{Ni}},
  \bibinfo{author}{\bibfnamefont{S.}~\bibnamefont{Ospelkaus}},
  \bibinfo{author}{\bibfnamefont{D.}~\bibnamefont{Wang}},
  \bibinfo{author}{\bibfnamefont{G.}~\bibnamefont{Qu{\'e}m{\'e}ner}},
  \bibinfo{author}{\bibfnamefont{B.}~\bibnamefont{Neyenhuis}},
  \bibinfo{author}{\bibfnamefont{M.~H.~G.} \bibnamefont{de~Miranda}},
  \bibinfo{author}{\bibfnamefont{J.~L.} \bibnamefont{Bohn}},
  \bibinfo{author}{\bibfnamefont{J.}~\bibnamefont{Ye}}, \bibnamefont{and}
  \bibinfo{author}{\bibfnamefont{D.~S.} \bibnamefont{Jin}},
  \bibinfo{journal}{Nature} \textbf{\bibinfo{volume}{464}},
  \bibinfo{pages}{1324} (\bibinfo{year}{2010}).

\bibitem[{\citenamefont{Ospelkaus et~al.}(2010)\citenamefont{Ospelkaus, Ni,
  Wang, de~Miranda, Neyenhuis, Qu\'{e}m\'{e}ner, Julienne, Bohn, Jin, and
  Ye}}]{Ospelkaus2010}
\bibinfo{author}{\bibfnamefont{S.}~\bibnamefont{Ospelkaus}},
  \bibinfo{author}{\bibfnamefont{K.-K.} \bibnamefont{Ni}},
  \bibinfo{author}{\bibfnamefont{D.}~\bibnamefont{Wang}},
  \bibinfo{author}{\bibfnamefont{M.~H.~G.} \bibnamefont{de~Miranda}},
  \bibinfo{author}{\bibfnamefont{B.}~\bibnamefont{Neyenhuis}},
  \bibinfo{author}{\bibfnamefont{G.}~\bibnamefont{Qu\'{e}m\'{e}ner}},
  \bibinfo{author}{\bibfnamefont{P.~S.} \bibnamefont{Julienne}},
  \bibinfo{author}{\bibfnamefont{J.~L.} \bibnamefont{Bohn}},
  \bibinfo{author}{\bibfnamefont{D.~S.} \bibnamefont{Jin}}, \bibnamefont{and}
  \bibinfo{author}{\bibfnamefont{J.}~\bibnamefont{Ye}},
  \bibinfo{journal}{Science} \textbf{\bibinfo{volume}{327}},
  \bibinfo{pages}{853} (\bibinfo{year}{2010}).

\bibitem[{\citenamefont{Micheli et~al.}(2010)\citenamefont{Micheli, Idziaszek,
  Pupillo, Baranov, Zoller, and Julienne}}]{Micheli2010}
\bibinfo{author}{\bibfnamefont{A.}~\bibnamefont{Micheli}},
  \bibinfo{author}{\bibfnamefont{Z.}~\bibnamefont{Idziaszek}},
  \bibinfo{author}{\bibfnamefont{G.}~\bibnamefont{Pupillo}},
  \bibinfo{author}{\bibfnamefont{M.~A.} \bibnamefont{Baranov}},
  \bibinfo{author}{\bibfnamefont{P.}~\bibnamefont{Zoller}}, \bibnamefont{and}
  \bibinfo{author}{\bibfnamefont{P.~S.} \bibnamefont{Julienne}},
  \bibinfo{journal}{Phys. Rev. Lett.} \textbf{\bibinfo{volume}{105}},
  \bibinfo{pages}{073202} (\bibinfo{year}{2010}).

\bibitem[{\citenamefont{Qu\'em\'ener and Bohn}(2010)}]{Quemener2010}
\bibinfo{author}{\bibfnamefont{G.}~\bibnamefont{Qu\'em\'ener}}
  \bibnamefont{and} \bibinfo{author}{\bibfnamefont{J.~L.} \bibnamefont{Bohn}},
  \bibinfo{journal}{Phys. Rev. A} \textbf{\bibinfo{volume}{81}},
  \bibinfo{pages}{060701} (\bibinfo{year}{2010}).

\bibitem[{\citenamefont{Qu\'em\'ener and Bohn}(2011)}]{Quemener2011}
\bibinfo{author}{\bibfnamefont{G.}~\bibnamefont{Qu\'em\'ener}}
  \bibnamefont{and} \bibinfo{author}{\bibfnamefont{J.~L.} \bibnamefont{Bohn}},
  \bibinfo{journal}{Phys. Rev. A} \textbf{\bibinfo{volume}{83}},
  \bibinfo{pages}{012705} (\bibinfo{year}{2011}).

\bibitem[{\citenamefont{Julienne et~al.}(2011)\citenamefont{Julienne, Hanna,
  and Idziaszek}}]{Julienne2011}
\bibinfo{author}{\bibfnamefont{P.~S.} \bibnamefont{Julienne}},
  \bibinfo{author}{\bibfnamefont{T.~M.} \bibnamefont{Hanna}}, \bibnamefont{and}
  \bibinfo{author}{\bibfnamefont{Z.}~\bibnamefont{Idziaszek}},
  \bibinfo{journal}{Physical Chemistry Chemical Physics}
  \textbf{\bibinfo{volume}{13}}, \bibinfo{pages}{19114} (\bibinfo{year}{2011}).

\bibitem[{\citenamefont{Zhu et~al.}(2013)\citenamefont{Zhu, Qu\'em\'ener, Rey,
  and Holland}}]{Zhu2013}
\bibinfo{author}{\bibfnamefont{B.}~\bibnamefont{Zhu}},
  \bibinfo{author}{\bibfnamefont{G.}~\bibnamefont{Qu\'em\'ener}},
  \bibinfo{author}{\bibfnamefont{A.~M.} \bibnamefont{Rey}}, \bibnamefont{and}
  \bibinfo{author}{\bibfnamefont{M.~J.} \bibnamefont{Holland}},
  \bibinfo{journal}{Phys. Rev. A} \textbf{\bibinfo{volume}{88}},
  \bibinfo{pages}{063405} (\bibinfo{year}{2013}).

\bibitem[{\citenamefont{Simoni et~al.}(2015)\citenamefont{Simoni, Srinivasan,
  Launay, Jachymski, Idziaszek, and Julienne}}]{simoni2015}
\bibinfo{author}{\bibfnamefont{A.}~\bibnamefont{Simoni}},
  \bibinfo{author}{\bibfnamefont{S.}~\bibnamefont{Srinivasan}},
  \bibinfo{author}{\bibfnamefont{J.-M.} \bibnamefont{Launay}},
  \bibinfo{author}{\bibfnamefont{K.}~\bibnamefont{Jachymski}},
  \bibinfo{author}{\bibfnamefont{Z.}~\bibnamefont{Idziaszek}},
  \bibnamefont{and} \bibinfo{author}{\bibfnamefont{P.~S.}
  \bibnamefont{Julienne}}, \bibinfo{journal}{New Journal of Physics}
  \textbf{\bibinfo{volume}{17}}, \bibinfo{pages}{013020}
  (\bibinfo{year}{2015}).

\bibitem[{\citenamefont{Idziaszek and Julienne}(2010)}]{Idziaszek2010}
\bibinfo{author}{\bibfnamefont{Z.}~\bibnamefont{Idziaszek}} \bibnamefont{and}
  \bibinfo{author}{\bibfnamefont{P.~S.} \bibnamefont{Julienne}},
  \bibinfo{journal}{Phys. Rev. Lett.} \textbf{\bibinfo{volume}{104}},
  \bibinfo{pages}{113202} (\bibinfo{year}{2010}).

\bibitem[{\citenamefont{Idziaszek et~al.}(2010)\citenamefont{Idziaszek,
  Qu\'em\'ener, Bohn, and Julienne}}]{Idziaszek2010a}
\bibinfo{author}{\bibfnamefont{Z.}~\bibnamefont{Idziaszek}},
  \bibinfo{author}{\bibfnamefont{G.}~\bibnamefont{Qu\'em\'ener}},
  \bibinfo{author}{\bibfnamefont{J.~L.} \bibnamefont{Bohn}}, \bibnamefont{and}
  \bibinfo{author}{\bibfnamefont{P.~S.} \bibnamefont{Julienne}},
  \bibinfo{journal}{Phys. Rev. A} \textbf{\bibinfo{volume}{82}},
  \bibinfo{pages}{020703} (\bibinfo{year}{2010}).

\bibitem[{\citenamefont{Jachymski et~al.}(2013)\citenamefont{Jachymski, Krych,
  Julienne, and Idziaszek}}]{Jachymski2013}
\bibinfo{author}{\bibfnamefont{K.}~\bibnamefont{Jachymski}},
  \bibinfo{author}{\bibfnamefont{M.}~\bibnamefont{Krych}},
  \bibinfo{author}{\bibfnamefont{P.~S.} \bibnamefont{Julienne}},
  \bibnamefont{and}
  \bibinfo{author}{\bibfnamefont{Z.}~\bibnamefont{Idziaszek}},
  \bibinfo{journal}{Phys. Rev. Lett.} \textbf{\bibinfo{volume}{110}},
  \bibinfo{pages}{213202} (\bibinfo{year}{2013}).

\bibitem[{\citenamefont{Jachymski et~al.}(2014)\citenamefont{Jachymski, Krych,
  Julienne, and Idziaszek}}]{Jachymski2014}
\bibinfo{author}{\bibfnamefont{K.}~\bibnamefont{Jachymski}},
  \bibinfo{author}{\bibfnamefont{M.}~\bibnamefont{Krych}},
  \bibinfo{author}{\bibfnamefont{P.~S.} \bibnamefont{Julienne}},
  \bibnamefont{and}
  \bibinfo{author}{\bibfnamefont{Z.}~\bibnamefont{Idziaszek}},
  \bibinfo{journal}{Phys. Rev. A} \textbf{\bibinfo{volume}{90}},
  \bibinfo{pages}{042705} (\bibinfo{year}{2014}).

\bibitem[{\citenamefont{Mayle et~al.}(2012)\citenamefont{Mayle, Ruzic, and
  Bohn}}]{Mayle2012}
\bibinfo{author}{\bibfnamefont{M.}~\bibnamefont{Mayle}},
  \bibinfo{author}{\bibfnamefont{B.~P.} \bibnamefont{Ruzic}}, \bibnamefont{and}
  \bibinfo{author}{\bibfnamefont{J.~L.} \bibnamefont{Bohn}},
  \bibinfo{journal}{Phys. Rev. A} \textbf{\bibinfo{volume}{85}},
  \bibinfo{pages}{062712} (\bibinfo{year}{2012}).

\bibitem[{\citenamefont{Mayle et~al.}(2013)\citenamefont{Mayle, Qu\'em\'ener,
  Ruzic, and Bohn}}]{Mayle2013}
\bibinfo{author}{\bibfnamefont{M.}~\bibnamefont{Mayle}},
  \bibinfo{author}{\bibfnamefont{G.}~\bibnamefont{Qu\'em\'ener}},
  \bibinfo{author}{\bibfnamefont{B.~P.} \bibnamefont{Ruzic}}, \bibnamefont{and}
  \bibinfo{author}{\bibfnamefont{J.~L.} \bibnamefont{Bohn}},
  \bibinfo{journal}{Phys. Rev. A} \textbf{\bibinfo{volume}{87}},
  \bibinfo{pages}{012709} (\bibinfo{year}{2013}).

\bibitem[{\citenamefont{Qu\'em\'ener et~al.}(2005)\citenamefont{Qu\'em\'ener,
  Honvault, Launay, Sold\'an, Potter, and Hutson}}]{Quemener2005}
\bibinfo{author}{\bibfnamefont{G.}~\bibnamefont{Qu\'em\'ener}},
  \bibinfo{author}{\bibfnamefont{P.}~\bibnamefont{Honvault}},
  \bibinfo{author}{\bibfnamefont{J.-M.} \bibnamefont{Launay}},
  \bibinfo{author}{\bibfnamefont{P.}~\bibnamefont{Sold\'an}},
  \bibinfo{author}{\bibfnamefont{D.~E.} \bibnamefont{Potter}},
  \bibnamefont{and} \bibinfo{author}{\bibfnamefont{J.~M.}
  \bibnamefont{Hutson}}, \bibinfo{journal}{Phys. Rev. A}
  \textbf{\bibinfo{volume}{71}}, \bibinfo{pages}{032722}
  (\bibinfo{year}{2005}).

\bibitem[{\citenamefont{Qu\'em\'ener et~al.}(2007)\citenamefont{Qu\'em\'ener,
  Launay, and Honvault}}]{Quemener2007}
\bibinfo{author}{\bibfnamefont{G.}~\bibnamefont{Qu\'em\'ener}},
  \bibinfo{author}{\bibfnamefont{J.-M.} \bibnamefont{Launay}},
  \bibnamefont{and} \bibinfo{author}{\bibfnamefont{P.}~\bibnamefont{Honvault}},
  \bibinfo{journal}{Phys. Rev. A} \textbf{\bibinfo{volume}{75}},
  \bibinfo{pages}{050701} (\bibinfo{year}{2007}).

\bibitem[{\citenamefont{Danzl et~al.}(2010)\citenamefont{Danzl, Mark, Haller,
  Gustavsson, Hart, Aldegunde, Hutson, and Nagerl}}]{Danzl2010}
\bibinfo{author}{\bibfnamefont{J.~G.} \bibnamefont{Danzl}},
  \bibinfo{author}{\bibfnamefont{M.~J.} \bibnamefont{Mark}},
  \bibinfo{author}{\bibfnamefont{E.}~\bibnamefont{Haller}},
  \bibinfo{author}{\bibfnamefont{M.}~\bibnamefont{Gustavsson}},
  \bibinfo{author}{\bibfnamefont{R.}~\bibnamefont{Hart}},
  \bibinfo{author}{\bibfnamefont{J.}~\bibnamefont{Aldegunde}},
  \bibinfo{author}{\bibfnamefont{J.~M.} \bibnamefont{Hutson}},
  \bibnamefont{and} \bibinfo{author}{\bibfnamefont{H.-C.}
  \bibnamefont{Nagerl}}, \bibinfo{journal}{Nat. Phys.}
  \textbf{\bibinfo{volume}{6}}, \bibinfo{pages}{265} (\bibinfo{year}{2010}).

\bibitem[{\citenamefont{Yan et~al.}(2013)\citenamefont{Yan, Moses, Gadway,
  Covey, Hazzard, Rey, Jin, and Ye}}]{yan2013}
\bibinfo{author}{\bibfnamefont{B.}~\bibnamefont{Yan}},
  \bibinfo{author}{\bibfnamefont{S.~A.} \bibnamefont{Moses}},
  \bibinfo{author}{\bibfnamefont{B.}~\bibnamefont{Gadway}},
  \bibinfo{author}{\bibfnamefont{J.~P.} \bibnamefont{Covey}},
  \bibinfo{author}{\bibfnamefont{K.~R.} \bibnamefont{Hazzard}},
  \bibinfo{author}{\bibfnamefont{A.~M.} \bibnamefont{Rey}},
  \bibinfo{author}{\bibfnamefont{D.~S.} \bibnamefont{Jin}}, \bibnamefont{and}
  \bibinfo{author}{\bibfnamefont{J.}~\bibnamefont{Ye}},
  \bibinfo{journal}{Nature} \textbf{\bibinfo{volume}{501}},
  \bibinfo{pages}{521–525} (\bibinfo{year}{2013}).

\bibitem[{\citenamefont{Hazzard et~al.}(2014)\citenamefont{Hazzard, Gadway,
  Foss-Feig, Yan, Moses, Covey, Yao, Lukin, Ye, Jin et~al.}}]{Hazzard2014}
\bibinfo{author}{\bibfnamefont{K.~R.~A.} \bibnamefont{Hazzard}},
  \bibinfo{author}{\bibfnamefont{B.}~\bibnamefont{Gadway}},
  \bibinfo{author}{\bibfnamefont{M.}~\bibnamefont{Foss-Feig}},
  \bibinfo{author}{\bibfnamefont{B.}~\bibnamefont{Yan}},
  \bibinfo{author}{\bibfnamefont{S.~A.} \bibnamefont{Moses}},
  \bibinfo{author}{\bibfnamefont{J.~P.} \bibnamefont{Covey}},
  \bibinfo{author}{\bibfnamefont{N.~Y.} \bibnamefont{Yao}},
  \bibinfo{author}{\bibfnamefont{M.~D.} \bibnamefont{Lukin}},
  \bibinfo{author}{\bibfnamefont{J.}~\bibnamefont{Ye}},
  \bibinfo{author}{\bibfnamefont{D.~S.} \bibnamefont{Jin}},
  \bibnamefont{et~al.}, \bibinfo{journal}{Phys. Rev. Lett.}
  \textbf{\bibinfo{volume}{113}}, \bibinfo{pages}{195302}
  (\bibinfo{year}{2014}).

\bibitem[{\citenamefont{Lang et~al.}(2008)\citenamefont{Lang, Winkler, Strauss,
  Grimm, and Denschlag}}]{Rb2ground}
\bibinfo{author}{\bibfnamefont{F.}~\bibnamefont{Lang}},
  \bibinfo{author}{\bibfnamefont{K.}~\bibnamefont{Winkler}},
  \bibinfo{author}{\bibfnamefont{C.}~\bibnamefont{Strauss}},
  \bibinfo{author}{\bibfnamefont{R.}~\bibnamefont{Grimm}}, \bibnamefont{and}
  \bibinfo{author}{\bibfnamefont{J.~H.} \bibnamefont{Denschlag}},
  \bibinfo{journal}{Phys. Rev. Lett.} \textbf{\bibinfo{volume}{101}},
  \bibinfo{pages}{133005} (\bibinfo{year}{2008}).

\bibitem[{\citenamefont{Moses et~al.}(2015)\citenamefont{Moses, Covey,
  Miecnikowski, Yan, Gadway, Ye, and Jin}}]{Covey2015}
\bibinfo{author}{\bibfnamefont{S.~A.} \bibnamefont{Moses}},
  \bibinfo{author}{\bibfnamefont{J.~P.} \bibnamefont{Covey}},
  \bibinfo{author}{\bibfnamefont{M.~T.} \bibnamefont{Miecnikowski}},
  \bibinfo{author}{\bibfnamefont{B.}~\bibnamefont{Yan}},
  \bibinfo{author}{\bibfnamefont{B.}~\bibnamefont{Gadway}},
  \bibinfo{author}{\bibfnamefont{J.}~\bibnamefont{Ye}}, \bibnamefont{and}
  \bibinfo{author}{\bibfnamefont{D.~S.} \bibnamefont{Jin}},
  \bibinfo{journal}{eprint arXiv:1507.02377}  (\bibinfo{year}{2015}).

\bibitem[{\citenamefont{Anderlini et~al.}(2007)\citenamefont{Anderlini, Lee,
  Brown, Sebby-Strabley, Phillips, and Porto}}]{Anderlini2007}
\bibinfo{author}{\bibfnamefont{M.}~\bibnamefont{Anderlini}},
  \bibinfo{author}{\bibfnamefont{P.~J.} \bibnamefont{Lee}},
  \bibinfo{author}{\bibfnamefont{B.~L.} \bibnamefont{Brown}},
  \bibinfo{author}{\bibfnamefont{J.}~\bibnamefont{Sebby-Strabley}},
  \bibinfo{author}{\bibfnamefont{W.~D.} \bibnamefont{Phillips}},
  \bibnamefont{and} \bibinfo{author}{\bibfnamefont{J.~V.} \bibnamefont{Porto}},
  \bibinfo{journal}{Nature} \textbf{\bibinfo{volume}{448}},
  \bibinfo{pages}{452} (\bibinfo{year}{2007}).

\bibitem[{\citenamefont{Busch et~al.}(1998)\citenamefont{Busch, Englert,
  Rzazewski, and Wilkens}}]{busch}
\bibinfo{author}{\bibfnamefont{T.}~\bibnamefont{Busch}},
  \bibinfo{author}{\bibfnamefont{B.-G.} \bibnamefont{Englert}},
  \bibinfo{author}{\bibfnamefont{K.}~\bibnamefont{Rzazewski}},
  \bibnamefont{and} \bibinfo{author}{\bibfnamefont{M.}~\bibnamefont{Wilkens}},
  \bibinfo{journal}{{Found. Phys.}} \textbf{\bibinfo{volume}{28}},
  \bibinfo{pages}{549} (\bibinfo{year}{1998}).

\bibitem[{\citenamefont{Scolarici and Solombrino}(2006)}]{mohms_czechjp2006}
\bibinfo{author}{\bibfnamefont{G.}~\bibnamefont{Scolarici}} \bibnamefont{and}
  \bibinfo{author}{\bibfnamefont{L.}~\bibnamefont{Solombrino}},
  \bibinfo{journal}{{Czech. J. Phys.}} \textbf{\bibinfo{volume}{56}},
  \bibinfo{pages}{935} (\bibinfo{year}{2006}).

\bibitem[{\citenamefont{Faisal and Moloney}(1981)}]{non-Hermitian}
\bibinfo{author}{\bibfnamefont{F.~H.~M.} \bibnamefont{Faisal}}
  \bibnamefont{and} \bibinfo{author}{\bibfnamefont{J.~V.}
  \bibnamefont{Moloney}}, \bibinfo{journal}{{J. Phys. B -- At. Mol. }}
  \textbf{\bibinfo{volume}{14}}, \bibinfo{pages}{3603} (\bibinfo{year}{1981}).

\bibitem[{\citenamefont{Huang and Huang}(2004)}]{china2003}
\bibinfo{author}{\bibfnamefont{C.~F.} \bibnamefont{Huang}} \bibnamefont{and}
  \bibinfo{author}{\bibfnamefont{K.~N.} \bibnamefont{Huang}},
  \bibinfo{journal}{{C.J.P.}} \textbf{\bibinfo{volume}{42}},
  \bibinfo{pages}{221} (\bibinfo{year}{2004}).

\bibitem[{\citenamefont{Aikawa et~al.}(2010)\citenamefont{Aikawa, Akamatsu,
  Hayashi, Oasa, Kobayashi, Naidon, Kishimoto, Ueda, and Inouye}}]{Aikawa2010}
\bibinfo{author}{\bibfnamefont{K.}~\bibnamefont{Aikawa}},
  \bibinfo{author}{\bibfnamefont{D.}~\bibnamefont{Akamatsu}},
  \bibinfo{author}{\bibfnamefont{M.}~\bibnamefont{Hayashi}},
  \bibinfo{author}{\bibfnamefont{K.}~\bibnamefont{Oasa}},
  \bibinfo{author}{\bibfnamefont{J.}~\bibnamefont{Kobayashi}},
  \bibinfo{author}{\bibfnamefont{P.}~\bibnamefont{Naidon}},
  \bibinfo{author}{\bibfnamefont{T.}~\bibnamefont{Kishimoto}},
  \bibinfo{author}{\bibfnamefont{M.}~\bibnamefont{Ueda}}, \bibnamefont{and}
  \bibinfo{author}{\bibfnamefont{S.}~\bibnamefont{Inouye}},
  \bibinfo{journal}{Phys. Rev. Lett.} \textbf{\bibinfo{volume}{105}},
  \bibinfo{pages}{203001} (\bibinfo{year}{2010}).

\bibitem[{\citenamefont{Deiglmayr et~al.}(2008)\citenamefont{Deiglmayr,
  Grochola, Repp, M\"ortlbauer, Gl\"uck, Lange, Dulieu, Wester, and
  Weidem\"uller}}]{Deiglmayr2008}
\bibinfo{author}{\bibfnamefont{J.}~\bibnamefont{Deiglmayr}},
  \bibinfo{author}{\bibfnamefont{A.}~\bibnamefont{Grochola}},
  \bibinfo{author}{\bibfnamefont{M.}~\bibnamefont{Repp}},
  \bibinfo{author}{\bibfnamefont{K.}~\bibnamefont{M\"ortlbauer}},
  \bibinfo{author}{\bibfnamefont{C.}~\bibnamefont{Gl\"uck}},
  \bibinfo{author}{\bibfnamefont{J.}~\bibnamefont{Lange}},
  \bibinfo{author}{\bibfnamefont{O.}~\bibnamefont{Dulieu}},
  \bibinfo{author}{\bibfnamefont{R.}~\bibnamefont{Wester}}, \bibnamefont{and}
  \bibinfo{author}{\bibfnamefont{M.}~\bibnamefont{Weidem\"uller}},
  \bibinfo{journal}{Phys. Rev. Lett.} \textbf{\bibinfo{volume}{101}},
  \bibinfo{pages}{133004} (\bibinfo{year}{2008}),
  \urlprefix\url{http://link.aps.org/doi/10.1103/PhysRevLett.101.133004}.

\end{thebibliography}

\end{document}